\documentclass{article}
 
\begin{document}
\title{ON THE COSMOLOGY OF WEYL'S GAUGE INVARIANT GRAVITY
}
\author{Takuya Maki\\
Japan Women's College of Physical Education, Nikaido University\\
	 Setagaya,~Tokyo 157-8565,~Japan
\\
\\
Yuji Naramoto
and
Kiyoshi Shiraishi\\
Faculty of Science, Yamaguchi University\\ 
	 Yoshida,Yamaguchi-shi,~Yamaguchi 753-8512,~Japan
}
\maketitle
\begin{abstract}
Recently the vector inflation has been proposed as the alternative
to inflationary models based on scalar bosons and quintessence scalar
fields. In the vector inflationary model, the vector field non-minimally
couples to gravity.  We should, however, inquire if there exists a
relevant fundamental theory which supports the inflationary scenario.
We investigate the possibility that Weyl's gauge gravity theory could be
such a fundamental theory. That is the reason why the Weyl's gauge
invariant vector and scalar fields are naturally introduced. After
rescaling the Weyl's gauge invariant Lagrangian to the Einstein frame, we
find that in four dimensions the Lagrangian is equivalent to
Einstein-Proca theory and does not have the vector field non-minimally
coupled to gravity,  but has the scalar boson with a polynomial potential
 which leads to the spontaneously symmetry breakdown. 
\end{abstract}

\section{Introduction}
Inflationary models are proposed as some resolutions for the cosmological
problems,  {\it e.g.}, the flatness, horizon and monopole problems.
These successful models, for example, chaotic inflation~\cite{L},
$k$-inflation~\cite{PDM}, are based on models of scalar bosons.
The chaotic inflationary model have at least a difficulty
in which bosonic fields could condensate some domains, {\it i.e.}, in the
early stage, some domains successfully exit  but others keep expanding.
In $k$-inflation and the modified modes~\cite{BM}, these inflation are
driven by non-minimal and non-canonical kinetic terms,
but need  some adjustments of conditions of scalar fields and its
potentials.
In addition we have detected no such scalar bosons by experiments. 
From these reasons, recently the vector inflation has been proposed 
by Ford~\cite{F} and some authors have studied the model~\cite{BL,GMV,GV}.
They consider the following action:
\begin{equation}
S=\int_{}^{} {d^4x}\sqrt {-g}\left\{
{\frac{R}{16\pi}-\frac{1}{4}F^2-\frac{1}{2} ( m^2+\frac{R}{6} ) A^2
 } \right\}\,, \label{action}
\end{equation} 
where $R$ is the scalar curvature, $F$ denotes the field strength of vector field 
$A
$ and $m$ is 
the mass of the vector field. It is worth noting that the massive vector
field non-minimally couples to gravity in Eq.~(\ref{action}). The
isotropy and the stability of the inflationary model have been
discussed~\cite{GMV}. The isotropy of expansion is achieved by {\it N}
randomly organized vector fields
 or by a triplet of  orthogonal vector fields. However, these discussions
have been made to solve the cosmological problems
 from a aspect of cosmological observations. 
These models are assumed the bosonic inflatons with  potentials
 that are not completely supported by fundamental physics. It is,
therefore, necessary to investigate how the fundamental physical theories
support the inflationary models.  

In the very early stage of our universe, the gravitational theory is
expected to be different from the ordinary Einstein gravity~\cite{CM},
{\it e.g.} higher derivative gravity, scalar-tensor gravity. Indeed,
quantum gravity or string corrections would affect the cosmological
evolution near the Planck scale. In particular, gravitational theory
could be speculated to be a scale invariant  in this stage as well as
other fundamental physics.

In this paper, we study the possibility that the Weyl's gauge gravity
is a such fundamental theory. 
Weyl's gauge theory of gravity is an extension of the Einstein
gravity~[9-24]. Especially  the vector and scalar
bosons are naturally introduced in this theory by the scale invariant
symmetry. We consider that 
Weyl's gauge invariant scalar and vector field are expected to play
cosmological important role in the very early stage of our universe.  In
Sec.~2, Weyl's gauge transformation is introduced as the local scale
transformation.
Then we construct the minimal Weyl's gauge invariant Lagrangian in
arbitrary  dimensions in Sec.~3.  In Sec.~4, we discuss the cosmology by
using the Lagrangian obtained in Sec.~3.

\section{Weyl's gauge gravity theory}
In this section, we review the Weyl's gauge transformation
to construct the gauge invariant Lagrangian.

Consider the scale transformation in $D$-dimensions
\begin{equation}
ds\rightarrow ds'=e^{\Lambda(x)}ds\,,
\end{equation}
where $\Lambda(x)$ is an arbitrary function of the coordinates 
$x^\mu$($\mu=0,\dots , D-1$).
Then the transformation of metric is realized by
\begin{equation}
g_{\mu\nu}\rightarrow g'_{\mu\nu}=e^{2\Lambda(x)}g_{\mu\nu}\,.
\label{trans_metric1}
\end{equation}
Thus
\begin{equation}
g^{\mu\nu}\rightarrow g'^{\mu\nu}=e^{-2\Lambda(x)}g^{\mu\nu}\,,
\end{equation}
and
\begin{equation}
\sqrt{-g}\rightarrow \sqrt{-g^\prime}=e^{D\Lambda(x)}\sqrt{-g}\,.
\end{equation}
We can define the field with weight $d=-\frac{D-2}{2}$ which transforms as
\begin{equation}
\Phi\rightarrow \Phi'=e^{-\frac{D-2}{2}\Lambda(x)}\Phi\,.
\end{equation}
We consider the covariant derivative of the scalar field 
\begin{equation}
\partial_{\mu}\Phi\Rightarrow
\tilde\partial_\mu\Phi\equiv
\partial_\mu\Phi-\frac{D-2}{2}\,A_\mu\Phi\,,
\end{equation}
where $A_\mu$ is a Weyl's gauge invariant vector meson and its field 
strength is given by
\begin{equation}
F_{\mu\nu}\equiv\partial_\mu A_\nu-\partial_\nu A_\mu\,.
\end{equation}
Under the Weyl's gauge field transformation
\begin{equation}
A_\mu\rightarrow A'_\mu=
A_\mu-\partial_\mu\Lambda(x)\,,
\label{gaugetrans}
\end{equation}
we obtain the transformation of the covariant derivative of the scalar field as
\begin{equation}
\tilde\partial_\mu\Phi\rightarrow
e^{-\frac{D-2}{2}\Lambda(x)}\tilde\partial_\mu\Phi\,.
\end{equation}
Moreover it is easily seen that
\begin{equation}
F_{\mu\nu}\rightarrow F'_{\mu\nu}=F_{\mu\nu}\,.
\end{equation}
The modified Christoffel symbol is defined as
\begin{equation}
\tilde\Gamma^\lambda_{\mu\nu}\equiv\frac{1}{2}g^{\lambda\sigma}
\left(\tilde\partial_\mu
g_{\sigma\nu}+\tilde\partial_\nu
g_{\mu\sigma}-\tilde\partial_\sigma
g_{\mu\nu}\right)\,,
\end{equation}
and the modified curvature is given as follows:
\begin{equation}
\tilde R^\mu{}_{\nu\rho\sigma}\equiv
\partial_\rho\tilde\Gamma^\mu_{\nu\sigma}-
\partial_\sigma\tilde\Gamma^\mu_{\nu\rho}+
\tilde\Gamma^\mu_{\lambda\rho}\tilde\Gamma^\lambda_{\nu\sigma}-
\tilde\Gamma^\mu_{\lambda\sigma}\tilde\Gamma^\lambda_{\nu\rho}\,.
\end{equation}
In Weyl's gauge theory of gravity, the Lagrangian should be invariant under the scale
transformation.

\section{Weyl invariant Lagrangian}
First, we show the Weyl's gauge invariant sectors of the vector field, 
the scalar field,
 the curvature $R$ and $R^2$ in $D$-dimensions:
\begin{eqnarray}
{\cal
L}_A&=&-\frac{1}{4e^2}\sqrt{-g}\,\Phi^{\frac{2(D-4)}{D-2}}g^{\mu\rho}g^{\nu\sigma}
F_{\mu\nu}F_{\rho\sigma}\,,\\
{\cal
L}_{\Phi}&=&-\sqrt{-g}\left[\frac{1}{2}g^{\mu\nu}{\tilde\partial}_{\mu}\Phi
{\tilde\partial}_{\nu}\Phi+\frac{1}{4}\lambda\Phi^{\frac{2D}{D-2}}\right]\,
,\\
{\cal
L}_{R}&=&\frac{1}{2}\sqrt{-g}\,\epsilon \Phi^{2}\tilde R\,,\\
{\cal
L}_{R^2}&=&\sqrt{-g}\,\alpha \Phi^{\frac{2(D-4)}{D-2}}{\tilde R}^2\,,
\end{eqnarray}
where $\lambda$, $\epsilon$, $e$ and $\alpha$ are dimensionless constants
and
\begin{equation}
 \tilde R=R-2(D-1)\nabla_\mu A^\mu-(D-1)(D-2)A_\mu A^\mu.
\label{R}
\end{equation}
As seen from (\ref{R}),  ${\cal L}_{R^2}$ seems to include the term
of
$R A_{\mu}A^{\mu}$.

The simple Lagrangian which consists of $A_\mu$,$\Phi$, $\tilde R$ and
$\tilde R^2$ is
 the combination of the above sectors.
Kao investigated the cosmology of Weyl's gauge gravity in four dimensions
\cite{Kao2}.
He focused on the higher derivative $R^{2}$ and introduced effective
scalar potentials.
Thus we take the more general higher derivative of $R$ into account,
then in general we consider the following Lagrangian
including higher order of the curvature $\tilde R^{n}$:
\begin{eqnarray}
{\cal
L}/\sqrt{-g}&=&-\frac{1}{4e^2}\,\Phi^{\frac{2(D-4)}{D-2}}
g^{\mu\rho}g^{\nu\sigma}
F_{\mu\nu}F_{\rho\sigma}-\frac{1}{2}g^{\mu\nu}{\tilde\partial}_{\mu}\Phi
{\tilde\partial}_{\nu}\Phi-\frac{1}{4}\lambda\Phi^{\frac{2D}{D-2}}\nonumber \\
&&+\frac{1}{2}\,\epsilon
\Phi^{\frac{2D}{D-2}}(\Phi^{\frac{-4}{D-2}}\tilde R)+\,\alpha
\Phi^{\frac{2D}{D-2}}(\Phi^{\frac{-4}{D-2}}\tilde R)^n.
\label{minimalaction}
\end{eqnarray}
Introducing an auxiliary field $\chi$, we get the equivalent Lagrangian as
\begin{eqnarray}
& &{\cal	
L}/\sqrt{-g}=-\frac{1}{4e^2}\,\Phi^{\frac{2(D-4)}{D-2}}
g^{\mu\rho}g^{\nu\sigma}
F_{\mu\nu}F_{\rho\sigma}-\frac{1}{2}g^{\mu\nu}{\tilde\partial}_{\mu}\Phi
{\tilde\partial}_{\nu}\Phi-\frac{1}{4}\lambda\Phi^{\frac{2D}{D-2}}
\qquad\quad\quad\nonumber
\\  & &+\frac{1}{2}\,\epsilon
\Phi^{\frac{2D}{D-2}}\chi+\,\alpha \Phi^{\frac{2D}{D-2}}\chi^n
+\left(\frac{1}{2}\,\epsilon
\Phi^{\frac{2D}{D-2}}+\,n\alpha
\Phi^{\frac{2D}{D-2}}\chi^{n-1}\right)\left(
\Phi^{\frac{-4}{D-2}}\tilde R-\chi\right).\label{Lagrangian1}
\end{eqnarray}
Furthermore the Lagrangian (\ref{Lagrangian1}) can be rewritten by the
new metric conformally related to the original one and new variables.
Here we choose
\begin{equation}
\hat{g}_{\mu\nu}\equiv e^{2\Lambda(x)}g_{\mu\nu}\,,
\end{equation}
and 
\begin{equation}
\hat{A}_\mu\equiv A_\mu-\partial_\mu\Lambda(x)\,,
\end{equation}
where
\begin{equation}
e^{-\Lambda(x)}=f\left(
\Phi^{2}+\,\frac{2n\alpha}{\epsilon}
\Phi^{2}\chi\right)^{-\frac{1}{D-2}}\,.
\end{equation}
Note that a mass scale $f$ was introduced here.

Now we can rewrite Eq.(\ref{Lagrangian1}) to the following Lagrangian
\begin{eqnarray*}
{\cal
L}/\sqrt{-\hat{g}}
&=&-\frac{1}{4e^2}\,
\phi^{2\frac{D-4}{D-2}}\hat{g}^{\mu\rho}\hat{g}^{\nu\sigma}
\hat{F}_{\mu\nu}\hat{F}_{\rho\sigma}-\frac{1}{2}
\left(\partial_\mu \phi-\frac{D-2}{2}\hat{A}_\mu \phi\right)^2\nonumber \\
&&-\frac{1}{4}
\lambda\,\phi^{\frac{2D}{D-2}}-(n-1)\alpha\left(\frac{\epsilon}{2n\alpha}
\right)^{\frac{n}{n-1}}
\phi^{\frac{2D}{D-2}-\frac{2n}{n-1}}
\left({f^{D-2}}-\phi^2\right)^{\frac{n}{n-1}}\nonumber \\
&&+\frac{1}{2}\,\epsilon f^{D-2}(\hat{R}-2(D-1)\hat{\nabla}_\mu
\hat{A}^\mu-(D-1)(D-2)\hat{A}_\mu \hat{A}^\mu )\,, 
\label{Lagrangian2}
\end{eqnarray*}
where
\begin{eqnarray}
\phi \equiv  f^{\frac{D-2}{2}} \left(1+\,\frac{2n\alpha}{\epsilon}
\chi^{n-1}\right)^{-1/2}\, ,
\end{eqnarray}
and `` $\hat{~}$ '' indicates the derived quantities from new variables.
We should note that  $\hat{R}\hat{A}_{\mu}\hat{A}^{\mu}$ term and
higher terms of the scalar curvature $\hat{R}$
 disappear in this expression.

\section{Cosmology of Weyl's gauge Gravity}
Since we obtained the Weyl's gauge invariant Lagrangian in the Einstein frame,
we can study the cosmology by using this Lagrangian. 
Also we consider in four dimensions: $D=4$ and the order of higher derivative of
curvature as $n=2$. 

The Lagrangian (\ref{Lagrangian2}) reads
\begin{eqnarray}
{\cal
L}/\sqrt{-\hat{g}}
&=&
\frac{1}{2}\,\epsilon f^{2}(\hat{R}
-6\hat{A}_\mu \hat{A}^\mu )
-\frac{1}{2}
\left(\partial_\mu\phi-\hat{A}_\mu \phi\right)^2 \nonumber \\
&&-\frac{1}{4e^2}\hat{F}^2
-\frac{\epsilon^2}{16\alpha}
\left({f^{2}}-\phi^2\right)^{2}
-\frac{1}{4}
\lambda\,\phi^{4}\,.
\label{4DminimalA}
\end{eqnarray}
This Lagrangian becomes markedly simple for $D=4$, namely, it consists of
a massive vector meson and a canonical scalar sector with a polynomial
potential which leads to a spontaneously symmetry breakdown. Hence, the
universe is expected to behave similarly to the well-known inflationary
scenario
 for our minimal Lagrangian (\ref{minimalaction}).
From Eq.(\ref{4DminimalA}), the conformal vector field could not be candidate of inflaton
 but could affect the evolution of our universe. 
The cosmology of Weyl's gauge gravity has been investigated in four dimensions 
 by Kao~\cite{Kao2}. 
 He has focused on the higher curvature term $R^{2}$
 and introduced an effective potential which the scalar field leads to symmetry-breakdown in the low energy region.
Also the vector mesons are not taken into account in contrast to our model. 
As this result, his model missed the canonical form of the scalar sector.

\section{Summary and Outlook}
In the early stage of the universe, vector inflations have been 
discussed as the alternative to the successfully inflationary scenarios
based on scalar bosons. However, these inflaton has not been completely
supported by relevant fundamental theory of physics. Also the gravity
theory is expected to be different from the Einstein gravity in the very
early universe. In particular, gravitational physics is speculated to
have a symmetry of 
 scale invariance near Planck scale like other particle physics.  

Thus, we study the possibility of the Weyl's gauge invariant theory as a
fundamental theory in the early universe.  One of the reasons is that the
Weyl's gauge invariant scalar and vector field can be naturally
introduced. We construct the Weyl's gauge invariant Lagrangian in
arbitrary dimensions that includes an arbitrary higher order  of the
scalar curvature  $R^n$.
This Lagrangian has the $RA^{\mu}A_{\mu}$ term.
In order to investigate the cosmology, we rewrite this Lagrangian to the
the Einstein-like form by using the Weyl's gauge transformation.

Especially, for $D=4$, the transformed Lagrangian is markedly simple.
In this Lagrangian, $\hat{R}$ is not minimally-coupled to the massive
vector $\hat{A}^{\mu}$. Therefore, the Weyl's gauge invariant vector
field could not be an inflaton of the vector inflation. However, the
Lagrangian has a scalar boson with polynomial potentials,  namely, the
canonical scalar sector with $\phi^4$-potential. Hence the universe
behaves similar to the ordinary one which has been discussed by many
authors.

While the massive vector field could not be inflaton of the vector
inflation,  nevertheless, it is expected that the vector meson relates to
the dark matter and dark energy.  In is worth noting that the study of
the cosmology of Weyl's gravity by Kao~\cite{Kao2}. In the contrast to
our model, he has  focused on the higher derivative $R^2$
and introduced  an effective scalar potentials but not taken the vector
fields into account.
From theas reasons, the effective action missed the canonical form of the
scalar sector.   From the Weyl's gauge gravity point of view,
if the inflaton is the Weyl's gauge invariant scalar,
the nature seems to select the polynomial potential instead of one in the
new inflation. 

We need to analyze the behavior of vector field
to obtain rigorous behavior of the inflaton.
It will be studied in a separate publication. Also we should investigate
the generalization to the case of higher and lower dimensions.
This will be published in the forthcoming paper.

\vspace{5mm}
This study is supported in part by the Grant-in-Aid of Nikaido Research
Fund.


\end{document}